\newcommand{\nc}{\newcommand}
\nc{\kms}{\,{km\,s$^{-1}$}}
\nc{\sgra}{Sgr\,A}
\nc{\sgrastar}{Sgr\,$\rm {A}^{*}$}
\nc{\sgraeast}{Sgr\,A\,East}
\nc{\sgrawest}{Sgr\,A\,West}
\nc{\sgracomp}{Sgr\,A\,complex}
\nc{\as}{\arcsec}
\nc{\HI}{H\,{\sc i}}
\nc{\HII}{H\,{\sc ii}}
\nc{\CI}{C\,{\sc i}}
\nc{\CII}{C\,{\sc i}}
\nc{\hto}{H$_{2}$O}
\nc{\htmo}{H$_{2}^{16}$O}
\nc{\htio}{H$_{2}^{18}$O}
\nc{\am}{\arcmin}
\nc{\ciso}{C$^{18}$O}
\nc{\hctn}{HC$_{3}$N}
\nc{\ot}{O$_{2}$}
\nc{\nht}{NH$_{3}$}
\nc{\htco}{H$_{2}$CO}
\nc{\htre}{H$_{3}$O${^+}$}
\nc{\ohs}{OH-streamer}
\nc{\s}{{$\mathrm{M}_{\odot} $}}
\nc{\snrs}{SNR (G359.92$-$0.09)}

\documentclass{aa}
\usepackage[varg]{txfonts}
\usepackage{graphicx}
\usepackage{txfonts}
\usepackage{appendix}

\let\textbf\relax

\begin{document}

\title{The OH-streamer in Sgr A  revisited: analysis of hydroxyl absorption within 10 pc from the Galactic centre. \thanks{Based on observations with the Karl G. Jansky Very Large Array, NRAO. The 1665 and 1667 MHz data cubes, presented in the on-line Appendices C and E, respectively, are also available in electronic form at the CDS (\textbf{at http://vizier.u-strasbg.fr/viz-bin/VizieR}).}}

\author{R.\,Karlsson\inst{1}
\and Aa.\,Sandqvist\inst{1}
\and K.\,Fathi\inst{1,2}
\and S.\,Mart\'in\inst{3}}

\institute{Stockholm Observatory, Stockholm University, AlbaNova
  University Center, SE-106 91 Stockholm, Sweden\\ \email{rolandk@astro.su.se}
\and Oskar Klein Centre for Cosmoparticle Physics, Stockholm University, SE-106 91 Stockholm, Sweden
\and Institut de RadioAstronomie Millim\'etrique 
300 rue de la Piscine, Domaine Universitaire 
38406 Saint Mart\'in d'H\`{e}res, France}

\offprints{\\ Roland Karlsson, \email{rolandk@astro.su.se}}

\date{Accepted for publication in A\&A}

\abstract{}
   {We study the structure and kinematics of the \ohs\ and
   the +80 \kms\ cloud and their interactions with the circumnuclear
   disk (CND) and with other molecular clouds in the vicinity of the
   Galactic centre (GC), and we map OH absorption at about 6\as\
   resolution at $R \leq$ 10 pc from the GC, with about 9 \kms\ of
   velocity resolution.} 
   {The VLA was used to map OH line absorption at the 1665 and 1667 MHz
   lambda doublet main lines of the $^{2}\Pi_{3/2}$ state towards the
   Sagittarius A complex.} 
   {Strong OH absorption was found in the \ohs, the southern streamer
   (SS), the +20, +50, and +80 \kms\ molecular clouds, the molecular
   belt, the CND, the expanding molecular ring (EMR), and the high
   negative velocity gas (HNVG). The \ohs\ was found to comprise
   three parts, head, middle, and tail, and to interact with the SS/+20,
   +80 \kms\ clouds and the CND. Optical depths and column densities 
\textbf{divided by excitation temperatures} have been calculated for the \ohs\ and the +80
   \kms\ cloud.}  {The \ohs, the SS, the +20 and +80 \kms\ clouds,
   and the CND are intimately related in position and velocity
   space. The \ohs\ was found to be a clumpy object stretching in
   projection from the inner radius of the CND at about 1.8 pc from
   \sgrastar towards and partly engulfing \sgrastar. As a side result of our
   data, a possible link between the near side of the EMR and the CND's
   southwest lobe was found. Additionally, we found OH absorption
   against all four of the previously known Compact \HII\ Regions A -
   D, located east of Sgr A East, indicating their close association
   with the +50 \kms\ cloud.}
  {}


\titlerunning{OH in GC molecular clouds}
\authorrunning{R. Karlsson et al.}
\maketitle

\section{Introduction}
At the very core of the Milky Way Galaxy is the $\sim$~4-million-solar-mass
supermassive black hole (SMBH) whose non-thermal radio continuum signature is
called \sgrastar. Surrounding it at a distance of one to a few pc and
orbiting it with a velocity of about 100 \kms is a rotating molecular
structure called the circumnuclear disk (CND). In the cavity inside the
CND, there is a central cluster of old stars and bright young stars
clustered around the SMBH. Inside the cavity is also the
mini-spiral-shaped \HII\ region called \sgrawest, which represents the inner
western edge of the CND, and a northern streamer extending from the CND
towards \sgrastar. A little farther out there is a large 
molecular belt extending from the southwest to the northeast (parallel
to the plane of the Galaxy). The major components of this belt are the
two giant molecular clouds, known as the +20 \kms\ cloud and the +50
\kms\ cloud. A giant energetic ($ > 10^{52}$ ergs) supernova-remnant-like non-thermal continuum radio 
shell (diameter about 8 pc), known as \sgraeast, is plowing into this
molecular belt from the side near \sgrastar, creating regions of shock
interaction especially at the inner surface of the +50 \kms\
cloud. Lunar occultation observations of OH absorptions at $\sim
40\as\ $ resolution showed the multiple structures within the Sgr A
region (Kerr \& Sandqvist \cite{ker68}; Sandqvist \cite{san73},
\cite{san74}). Interferometric observations of OH absorption in the
\sgracomp\ with an angular resolution of 4\farcm5 were subsequently
presented by Bieging \cite{bie76}. Absorption measurements are useful
for revealing the location of the clouds along the line of sight, relative
to the continuum emission.  

General reviews of the Galactic centre (GC) region have been presented by Mezger et al. (\cite{mez96}) and Morris \& Serabyn (\cite{mor96}), among others, and an
up-to-date introduction to the \sgracomp\ is given by Ferri\`ere
(\cite{fer12}).

We performed observations of the GC region in all four 18-cm OH
lines with the Karl G. Jansky Very Large Array, henceforth the VLA, in
the wide array BnA configuration and in the two main OH lines also in
the compact DnC configuration. The first preliminary 1667 MHz OH line
results with about 4\as\ angular resolution were presented by
Sandqvist et 
al. (\cite{san87}, \cite{sad89}). One of the most interesting results was the
discovery of an ``OH-streamer'' that stretches from the southwest part
of the CND and subsequently sweeps northeastward projected {\it \emph{inside}} the CND
towards \sgrastar. However, the velocities of the OH-streamer and the
SW CND differed by more than 100 \kms\, so their relation was
unclear. Subsequently, the \ohs\ was also detected at 1612 and 1665
MHz in the high-resolution BnA observations with the VLA (Karlsson et
al. \cite{kar03}). The full extent of the OH-streamer has not been
detected in other molecular lines, probably owing to a lack of
angular resolution and sensitivity. However, recent interferometric
observations of CN $J=2-1$ emission by Mart\'in et al. (\cite{mar12})
have revealed a few high-density clumps in the \ohs. No magnetic
field with a $3\sigma$ upper limit of 2 mG has been found in the
\ohs\ (Killeen et al. \cite{kil92}). Sandqvist et
al. (\cite{san87}) also briefly noted ``another feature at +78 \kms''
(henceforth the +80 \kms\ cloud) that seemed related to the \ohs.  

In this paper, we present the concatenated VLA BnA and DnC
observations of the two 1665 and 1667 MHz main OH lines. We
concentrate here on the interplay between the \ohs, the +80 \kms\
cloud, and the CND and possible effects of the SMBH and the surrounding
star cluster. Moreover, a possbile link between the CND SW lobe and
the near side of the expanding molecular ring is suggested. A detailed
study of all the features seen in the data and their interrelations
is, however, beyond the present scope and will be the subject of a
subsequent paper.  

\section{Observations}

\begin{figure}[h]
\begin{center}
\includegraphics[angle=0, width=.49\textwidth]{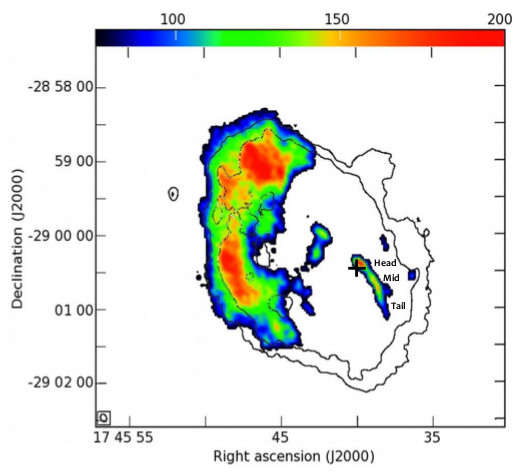}
\caption{OH absorption, (continuum minus line flux), at 1667 MHz
  shown in colour shades at a velocity of 59 \kms. The contour lines
  depict the 18 cm continuum emission of the \sgracomp\ at intensities
  of 100 and 150 mJy/beam. indicated with a plus sign. The \ohs\ is seen running in
  a northeasterly direction and ends slightly northwest of the
  position of Sgr A*, and its parts are labelled in the figure. The
  +50 \kms\ cloud is seen E of \sgrastar. } 
\label{OHSCOL2000_1.pdf}
\end{center}
\end{figure}

The main OH transition lines at 1665 and 1667 MHz and the satellite
lines at 1612 and 1720 MHz were observed in June 1986 with the VLA in
the BnA wide array configuration. The main transition lines were also
observed with the VLA in its DnC compact array configuration in
October 1989. The initial- and post-calibration of the observational
data were done with the National Radio Astronomy Observatory (NRAO),
Astronomical Image Processing \textbf{System} (AIPS) at the VLA site
and at the Array Operations Center in Socorro, NM. The two sets of
($u,v$) data for each of the main lines were subsequently concatenated
into one data set to improve sensitivity and reduce the effects of missing
zero-spacing observations. The concatenation was done with the NRAO
AIPS DBCON program package, and baselines were subtracted with the
NRAO AIPS UVLSF program package in the ($u,v$)-plane. Throughout
this paper we define ``OH absorption'' as ``continuum minus
line flux'' ($=-T_{\rm L}$), in units of mJy/beam. The angular
resolution is $7\as\times 5\as$, and the velocity resolution is 8.8
\kms. Typical noise levels in the concatenated spectral line maps and
line profiles are about 25 mJy/beam, and about 5 mJy/beam for the
original $4\farcs0 \times 2\farcs8$ angular resolution BnA 1667 MHz
data. The concatenated data are used for the main part of this paper,
but the original BnA 1667 MHz data are used as a complement. The
subsequent processing of the data was done with both the NRAO AIPS and
the XS\footnote{Developed by P. Bergman, Onsala Space Observatory}
program packages. Projected distances correspond to about 0.04 pc/\as\
assuming a distance to the GC of 8 kpc. The position of \sgrastar\ is
(RA, Dec) = 17$^{\rm h}$45$^{\rm m}$40\fs03, $-29\degr$$00'$27\farcs6
(J2000.0). For further information about the observations, see
Karlsson et al. (\cite{kar03}, \cite{kar13}).

\section{Results}

\begin{figure}[h]
\begin{flushleft}
\includegraphics[angle=0, width=.49\textwidth]{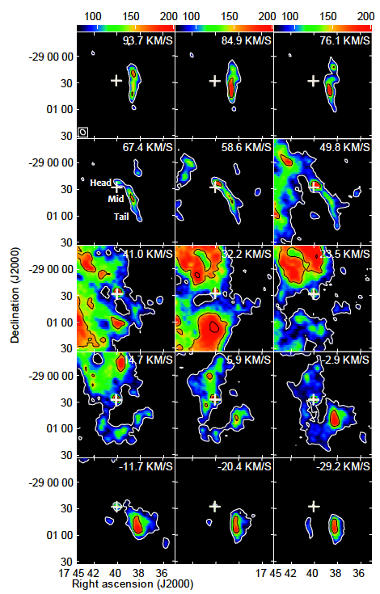}
\caption{OH absorption at 1667 MHz towards the inner $\sim 110$\as\
  ($\sim4.4$ pc) of the GC at velocities between $-$ 29 and 94
  \kms. The lowest contour level is 75 mJy/beam, $\sim3$$\sigma$, and
  the contour spacing is $\sim3$$\sigma$. The position of Sgr A* is
  marked with a plus sign. The \ohs\ is seen at velocities between
  $-$29 and 67 \kms, the +80 \kms\ cloud at velocities from 76 to 94
  \kms, and the CND SW lobe between $-$29 and 15 \kms. The three
  parts of the \ohs\ are labelled in the 67 \kms\ panel. The SS and
  the +20 \kms\ cloud can be seen S of \sgrastar\ in the 24 to 41
  \kms\ panels, and are most prominent at 32 \kms. The beam size is
  shown in the lower left corner of the uppermost left panel.} 
\label{OHSPCOL2000_2.pdf}
\end{flushleft}
\end{figure}

\subsection{Morphology and physical properties of the \ohs\ and the +80 \kms\ cloud}

The \ohs\ is observed at 1667 MHz between velocities of $-$29 and 67
\kms\ (Figs. \ref{OHSCOL2000_1.pdf} and \ref{OHSPCOL2000_2.pdf}; and in
the on-line-only Appendice Figs. B.1- B.3, C.1 - C.3). At 1665 MHz it
is observed between $-$20 and 68 \kms\ (Figs. D.1 and E.1). It is slightly curved to the E
with an average position angle (PA) of $\sim28\degr$ east of the north
celestial pole. The projected length of the \ohs\ is $\sim$
68\as\ (2.7 pc), and the width is $\sim$ (4 - 15)\as\ (0.2 - 0.6 pc),
as measured at 59 \kms. Three clumps, head, mid, and tail, are seen
inside of the \ohs\ at velocities of 50, 59, and 67 \kms in both the
1665 and 1667 MHz transitions. Inside of the mid part, a fine structure
of smaller clumps is also observed.  

\begin{figure}[h]
\includegraphics[angle=0, width=.49\textwidth]{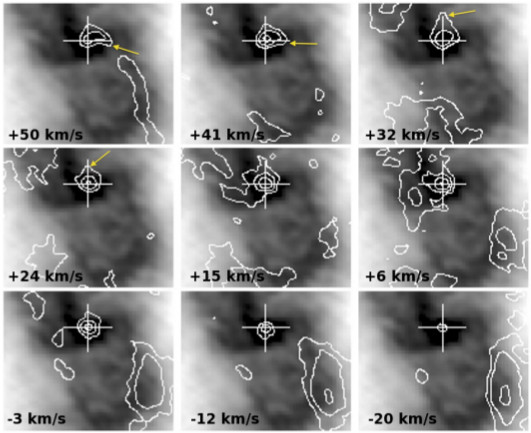}
\caption{OH absorption of the \ohs\ Head at 1667 MHz, observed with
  the VLA in BnA configuration at an angular resolution of $4.0\as
  \times 2.8 \as$ (Karlsson et al. \cite{kar03}) overlaid on the 18 cm
  continuum. We note the location of the central parts of Sgr A West in
  dark shades of grey. \sgrastar\ is labelled with a plus sign, and each
  panel covers a region of about $45\as \times  45\as$. The lowest
  contour level corresponds to about 4$\sigma$, and the velocity is
  given in the left-hand lower corner in each panel. We note the winglike
  appearance of the Head, marked by an arrow, and its anti-clockwise
  rotation as the velocity decreases from 50 to 24 \kms.} 
\label{headBnA}
\end{figure}

The geometry of OH-absorption in the head varies significantly with
velocity, as can be seen in Fig. \ref{headBnA}, where we present the higher
resolution BnA maps from Karlsson et al. (\cite{kar03}). A winglike structure develops on the western side
at velocities of about 50 \kms and displays an anti-clockwise
rotation as the velocity decreases, and the head becomes nearly
circular at velocities below 15 \kms. 

The +80 \kms\ cloud is observed at a distance of about 18\as\ (0.7 pc)
W of \sgrastar\ at velocities between 67 and 111 \kms\
(Figs. \ref{OHSPCOL2000_2.pdf}, \ref{OHSCND802000_inv.pdf} and B.1,
C.1). It is extended mainly in the north - south direction. The length extends
to 65\as\ (2.6 pc), while the width is about 15\as\ (0.6 pc) at 76
\kms. Two clumps are seen in the +80 \kms\ cloud between 76 \kms\ and
103 \kms, separated by $\sim25\as$ (1 pc). At 103 and 111 \kms, the
northern part of the cloud bends to the east, and points towards the
northeastern extension of the CND northeastern lobe in a similar manner to
a CN emitting feature, centred at 90 \kms\ in a 30 \kms\ wide bin
(Mart\'in et al. \cite{mar12}). 

By dividing the spectral line maps ($-T_{\rm{L}}$) by the continuum
map $(T_{\rm{C}}$), we reduce the effects of the varying continuum
intensity in the region and obtain ($-$$T_{\rm{L}}/T_{\rm{C}}$) maps
at 1667 MHz and 1665 MHz. This ratio is proportional to the optical
depth as $-T_{\rm{L}}/T_{\rm{C}}=1- \rm{e}^{-\tau}$, where 1 $-$
$\rm{e}^{-\tau}$ is often referred to as ``apparent opacity". Both the
\ohs\ and the +80 \kms\ cloud retain their basic properties in Figs. C.1 and C.2, which validates the reality of the objects. 

OH absorption line profiles were produced at 16 positions
centred along the \ohs\ and the +80 \kms\ cloud, respectively (Figs. A.1
and A.2.).
Optical depths ($\tau_{1667}$) and column densities divided by excitation 
temperatures ($N$(OH)/$T$$_{\rm{ex}}$) were calculated at the positions of 
the profiles as described in Karlsson et al. (\cite{kar13}) (see Table A.1).

To get a first order of the mass of
the \ohs\ and the +80 \kms\ cloud, we assume that each of the three
clumps in the \ohs and the full length of the +80 \kms\ cloud can be
represented by prolate ellipsoids (see Table \ref{table:ohsvirial}). The material between the clumps in the
\ohs\ are not considered, such that the mass of the \ohs\ is
considered as a lower limit. The value of $n_{\rm H_{2}}$ is taken as
$10^{5}$ $\rm cm^{-3}$, which is in the lower range for molecular gas
in the CND found in Ferri\`ere
(\cite{fer12}). We derive at a total mass of $\gtrsim$ 400 M$_\odot$ for
the OH-streamer and about 2500 M$_\odot$ for the more massive +80
\kms\ cloud (Table \ref{table:ohsvirial}).

\begin{table}
\caption{Properties of the \ohs\ and the +80 \kms\ cloud (assuming $n_{\rm H_{2}}$=$10^5$ $\rm{cm}^{-3}$).}
\begin{flushleft}
\scalebox{1}{
\begin{tabular}{lccccc}
\hline\noalign{\smallskip}
Object &Length x Width&  $D$  \tablefootmark{a} & Volume & Mass 
\\ [0.5ex]
 & (\as\ x \as)(pc x pc) &(pc)  & ($\rm{pc}^{3}$)  &   (\s) \\ 
\hline\noalign{\smallskip}
\hline\noalign{\smallskip}
Head &(15 x 5) (0.6 x 0.2)& 0.3& 0.013& 65\\  
Mid & (20 x 5) (0.8 x 0.2)&0.9 & 0.017& 85\\  
Tail &  (15 x 10) (0.6 x 0.4)&1.5& 0.050 &250\\  
+80 \kms\ cl. & (65 x 15) (2.6 x 0.6)&0.7& 0.490 &2450 \\ 
\noalign{\smallskip}\hline
\end{tabular}
\label{table:ohsvirial}}
\end{flushleft}
\tablefoottext{a}{Projected distance from the position of \sgrastar.}
\end{table}

\subsection{Locations and interactions of the components}

We present here three different
displays of our data to shed some light on possible links between the
objects: {\em i)} maps of projected locations, {\em ii)} position-velocity diagrams, and {\em iii)} a position angle-velocity diagram.  

\begin{figure}
\begin{center}
\includegraphics[angle=0, width=.49\textwidth]{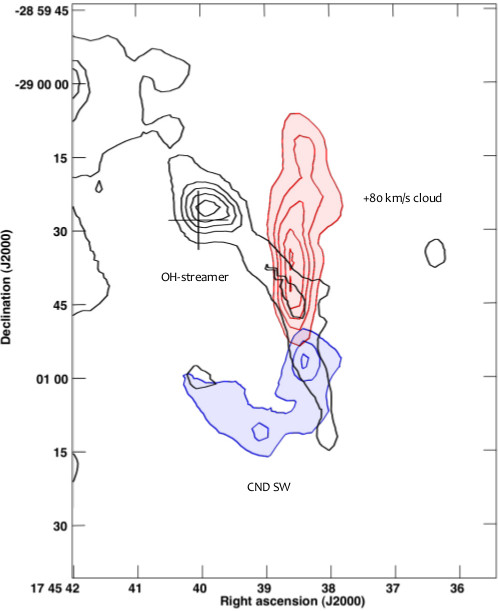}
\caption{OH absorption at 1667 MHz of the \ohs\ at 50 \kms\ with
  overlays of the +80 \kms\ cloud at 85 \kms, and the SW lobe of the
  CND at $-$73 \kms. The lowest contour level is drawn at $\sim90$
  mJy/beam ($\sim3.5$$\sigma$) and the contour spacing is
  $1\sigma$. \sgrastar\ is marked with a plus sign.}  
\label{OHSCND802000_inv.pdf}
\end{center}
\end{figure}

{\em i) Projected locations:} To study the relative
  positions of the OH-streamer, the +80 \kms\ cloud, and the CND
  southwest lobe, we have overlaid those three components in
  Fig. \ref{OHSCND802000_inv.pdf}. The \ohs\ head is observed in 
  absorption against \sgrastar, symmetrically between velocities of
  $-$29 to 15 \kms. At 24 to 67 \kms, the absorption moves to the
  northwest and is then only partly seen against \sgrastar\ (Figs. \ref{OHSPCOL2000_2.pdf}, \ref{headBnA}, and \ref{OHSCND802000_inv.pdf}). The Mid part overlaps the southern part of
  the +80 \kms\ cloud (Fig. \ref{OHSCND802000_inv.pdf}). The tail
  partially overlaps the CND southwestern lobe in
  Fig. \ref{OHSCND802000_inv.pdf}. In Fig. \ref{OHSPCOL2000_2.pdf} the northern
  part of the +20 \kms\ cloud and the SS are clearly seen about 60\as\
  (2.4 pc) south of the position of \sgrastar, at 32 \kms, which also
  is in the region of the southwest lobe of the CND. 

\begin{figure}[h]
\begin{center}
\includegraphics[angle=0, width=.48\textwidth]{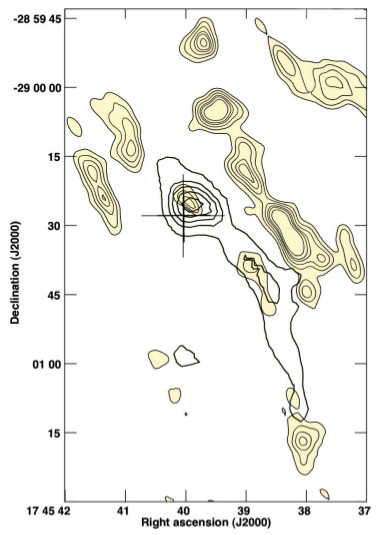}
\caption{Contour plot of CN $J=2-1$ integrated emission between 15 and
  45 \kms\ (shadowed areas) (Mart\'in et
  al. \cite{mar12}), with overlay of OH absorption at 50 \kms\ at 1667
  MHz. Three CN $J=2-1$ blobs fall inside, or
  partly inside, of the \ohs, and one is immediately S of the Tail. The
  lowest contour level of CN $J=2-1$ emission is $\sim3\sigma$, and $\sim4\sigma$ of
  OH. \sgrastar\ is marked with a plus sign. The CN emission and the head and mid parts of the \ohs\ are seemingly parallel to the Galactic plane.}  
\label{CNOHS_3.pdf}
\end{center}
\end{figure}

Three blobs of CN $J=2-1$ emission (Mart\'in
et al. \cite{mar12}) are observed in the \ohs\ head, mid, and tail
(Fig. \ref{CNOHS_3.pdf}). Moreover, $\rm C^{34}$S emission is also found in
the head, as seen in the lower left-hand panel of Fig. 6 in Liu et
al. (\cite{liu12}). Both the CN $J=2-1$ and $\rm C^{34}$S $J=7-6$
emission trace high-density regions. 

\begin{figure}[]
\begin{center}
\includegraphics[angle=0, width=0.49\textwidth]{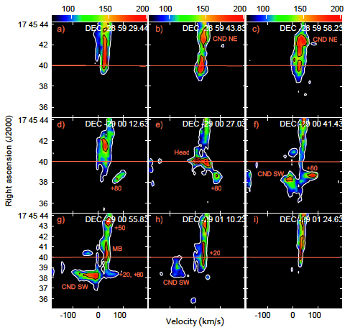}
\caption{Position-velocity diagrams (RA, Vel) of OH absorption at 1667
  MHz of an $\sim$2$\am \times 2\am$ region around \sgrastar. The
  lowest contour level is 75 mJy/beam ($\sim$$3\sigma$) and the
  contour spacing is also $\sim$$3\sigma$. The horizontal line
  indicates the right ascension of \sgrastar, and panel e) is drawn at
  the declination of \sgrastar. (``MB'' stands for molecular belt).}   
\label{RAPANEL2000_3.pdf}
\end{center}
\end{figure}

{\em ii) Position-velocity diagrams:} Position-velocity cuts of
OH absorption at 1667 MHz for a 2\am\ x 2\am\ region around
\sgrastar\ are shown in Fig. \ref{RAPANEL2000_3.pdf} and provide
a detailed picture of interactions in the position-velocity
space. Figure \ref{RAPANEL2000_3.pdf} e) is the RA scan at the
declination of \sgrastar\,, and it passes from the +80 \kms\ cloud to
the head of the \ohs. In the head, both negative and positive
velocities appear.  In Fig. \ref{RAPANEL2000_3.pdf} d), the +80
\kms\ cloud displays an increasing velocity from about 63 to 107
\kms in the easterly direction towards the CND NE extension,
corresponding to a velocity gradient of about 4.9 \kms /\as. A rapid
acceleration, from about 30 to 80 \kms\ is also observed in the
western part of the molecular belt/+20 \kms\ cloud, seen as the
knee-like structure at an (RA) of about 17$^{\rm h}$45$^{\rm m}$38\fs5 
in Figs. \ref{RAPANEL2000_3.pdf} f) and g). This occurs
in the overlap region of the \ohs\ Mid and the southern part of the
+80 \kms\ cloud. Moreover, the $\Delta V_{\rm FWHM}$ increases from 26 to 50 \kms\ in this region.

\begin{figure}[]
\begin{center}
\includegraphics[angle=0, width=.49\textwidth]{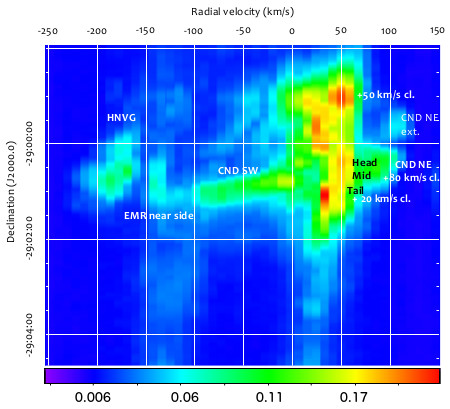}
\caption{Position-velocity diagram (Dec, Vel) of OH absorption at 1667
  MHz. This is a visualisation through the entire data cube
  as seen from the declination-velocity side. The 1665 MHz data overlap at
  velocities higher than about 160 \kms (see Fig. 3 in Sandqvist
  \cite{san73}). The wedge scale indicates the OH absorption in
  Jy/beam. (``HNVG'' stands for high negative-velocity gas).}  
\label{decvellarge_2000.pdf}
\end{center}
\end{figure}

 Figure \ref{decvellarge_2000.pdf} is a visualisation throughout the entire data cube, covering an $\sim 7\am\times 7\am\ $region seen face-on from the declination side. It illustrates that right ascension structure can also be easily followed at the same time. The corresponding visualisation seen from the right ascension-velocity side is shown in Fig. \ref{ravellarge_2.pdf}. 
The \ohs\ tail is seen to interact with the +20 \kms\ cloud, and
interactions between the +20, +80 \kms\ clouds can also be traced in
Fig. \ref{decvellarge_2000.pdf}. In those two figures the near side of the
EMR seems to connect to the CND SW lobe. The interaction occurs at
(RA, Dec) = 17$^{\rm h}$45$^{\rm m}$$39^{\rm s}$,
$-29\degr$$00'$$55^{\rm m}$, and the map position of this interaction
is located on the southeastern side of the \ohs\ tail. 

\begin{figure}[]
\includegraphics[angle=0, width=.49\textwidth]{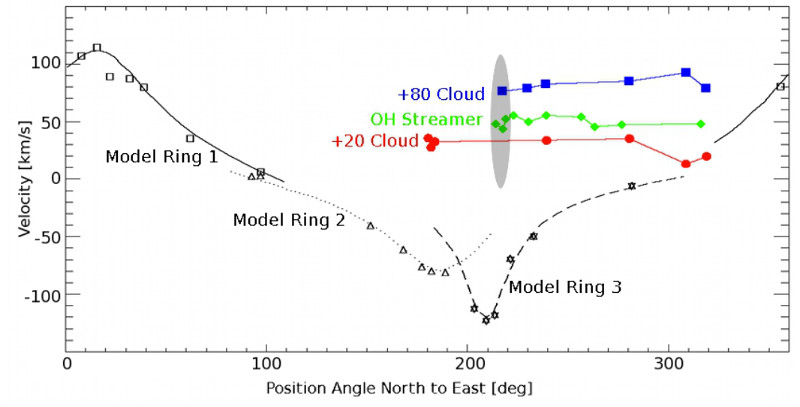}
\caption{Nominal velocities as a function of position angle E of N, as
  measured from the position of \sgrastar, of the \ohs\ (green filled
  diamonds), the +80 \kms\ cloud (blue filled squares), and the +20
  \kms\ cloud (red filled circles) overlaid on a
  diagram by Mart\'in et al. (\cite{mar12}), which shows the location
  of selected objects observed in CN emission. The ellipse indicates
  the region where interaction between the +20, +80 \kms\ clouds and
  the \ohs\ is suggested. The overlaid sinusoidal-like lines represent
  rotating ring models fitted to the core velocities in the southwest
  lobe (dashed line), southern extension (dotted line), and northeast
  lobe of the CND (solid line).} 
\label{Martin Fig 9 - 2 May.jpg}
\end{figure}

{\em iii) Position angle-velocity diagram:} Figure \ref{Martin Fig 9 - 2 May.jpg} 
is a modified diagram by Mart\'in et al. (\cite{mar12}), their Fig. 9, 
which shows measured component velocities in CN $J=2-1$ emission as 
a function of position angle (PA) measured from the position of \sgrastar. 
In this figure we have suppressed all points 
observed by Martin et al. (\cite{mar12}), except for those belonging 
to their rotating partial ring models for the CND, and superimposed 
our own calculated velocities and PAs for the \ohs\ and the +80 \kms\
cloud at Positions 1 - 16 (Fig. A.1 and Table \ref{table:paramsum}).

The \ohs\ shows a nearly flat velocity structure of about 50 \kms\ in the head
between PAs from about 260\degr\ to 310\degr\ in Fig. \ref{Martin Fig 9 - 2 May.jpg}. In the higher resolution BnA observations in Fig. \ref{headBnA}, we find that a winglike structure of the head occurs at 50 \kms and persists to $-$3 \kms, while displaying an anti-clockwise rotation (the arrows in Fig. \ref{headBnA}), and may be a signature of rotation of the OH gas in the head about \sgrastar\  or infalling of gas towards \sgrastar. The OH absorption also declines while the velocity drops from 50 \kms\ to $-$ 20 \kms. In Fig. \ref{Martin Fig 9 - 2 May.jpg}, the velocity is seen to increase about 10 \kms\
between PAs of about 230\degr\ and 260\degr\ corresponding to the mid
part of the \ohs, which is also the overlap region of the mid part and the +80
\kms\ cloud. The velocity of the two  outermost points in the tail decreases
towards the values in the +20 \kms\ cloud; this region is indicated with
an ellipse in Fig. \ref{Martin Fig 9 - 2 May.jpg}. 

The +80 \kms\ cloud is sampled well at PAs between
250$\degr$ and 315$\degr$ in Fig. \ref{Martin Fig 9 - 2
  May.jpg}, a similar interval as the ``70 \kms\ sources" observed
in CN emission (Mart\'in et al. \cite{mar12}), and it appears that
those sources are located in the +80 \kms\ cloud. It is clear from
this figure that the dominant part of the +80 \kms\ cloud is {\it \emph{not}} part
of the western edge of the CND. However, its northernmost part
might be connecting to the CND rotating ring model 1 in the
sector of PAs about 320\degr\ to 330\degr.

The +20 \kms\ cloud shows a flat velocity signature from PAs of
180\degr\ to 280\degr, after which the velocity appears to approach
that of the CND rotating ring model 3, near a PA of 310\degr.

\section{Concatenated OH data cubes}
Spectral line maps of OH absorption at 1667 MHz with an angular
resolution of $7\as \times\ 5\as$ are provided in Appendix B at
velocities between $-$196 and 111 \kms. The corresponding $-T_{\rm
  L}/T_C$ OH maps at 1667 MHz are presented in Appendix C. As a
representative of the 1665 MHz OH maps, we have chosen a few maps at
velocities between 15 and 59 \kms\, which are presented in Appendices D
and E. This limited velocity range covers some of the more pertinent
components in this paper.

In additon to the objects discussed in this paper, many other known
objects, such as the foreground spiral arm features at velocities of
about $-$50, $-$30, and 0 \kms, the high negative velocity gas, and the
wide velocity feature (Karlsson et al. \cite{kar13}), can be seen in
those maps.   

Although much additional information can be gleaned from the maps, we
wish here to restrict ourselves to an update concerning the Compact
\HII\ regions (A - D) (Ekers et al. \cite{eke83}) on the eastern edge
of the +50 \kms\ cloud. In Karlsson et al. (\cite{kar13}), we reported observations of OH absorption against the Compact \HII\ Region D. A
further analysis of the data now reveals that OH absorption is also
seen against the three other regions, A, B, and C, in the velocity
ranges of  24 to 76 \kms\ and 32 to 76 \kms, for the 1667 and 1665 MHz
lines, respectively (Figs. C.1, C.2, D.1, and E.1).

The four complete data cubes of the OH absorption and $-T_{\rm
  L}/T_C$ at 1665 and 1667 MHz are available in electronic form at the CDS
\textbf{at http://vizier.u-strasbg.fr/viz-bin/VizieR}.

\section{Discussion} 

The \ohs\ is clearly observed between velocities of $-$29 and 67 \kms. At velocities between 0 and 10 \kms\ confusion with foreground sources may occur. The \ohs\ is 
not observed in the $-T_{\rm{L}}/T_{\rm{C}}$ maps at velocities less than
41 \kms. Clipping of the different random noise levels in the channel
and continuum maps and absorption in foreground objects may explain
the absence of this feature at those velocities in the
$-T_{\rm{L}}/T_{\rm{C}}$ maps.  

Species like CN and OH tend to be abundant in  photo dissociation regions, 
suggesting that they mark the regions where UV radiation from the central 
star cluster interacts with the surrounding neutral gas. The
OH column densities divided by excitation temperatures
($N$(OH)/$T$$_{\rm{ex}}$) for the different parts are given in Table
\ref{table:paramsum}. It is likely that $T_{\rm ex}$ increases with
decreasing distance from \sgrastar, which would imply increasing OH column densities towards the head. 

In the region of the head, UV-radiation pressure from the
central star cluster may be a factor that could lead to
outflow of OH gas from the head. However, Murray et al. (\cite{mur11}) 
find that outflows from clusters in a Milky way-like galaxy (their
Fig. 4), would lead to outflow velocites of several hundreds of \kms,
at distances from the centre relevant for the \ohs\
head (0.3 pc), where we observe velocities of about 50 \kms. At
distances less than about 0.1 pc, the radiation pressure may, however,
be significant in supporting self-gravitational disks that are
supposed to fuel the SMBH at the centre (Thompson, \cite{tho09}). We
therefore argue that radiation pressure is not significant for the
kinematics of the head.  

The influence radius of the SMBH and the surrounding star cluster can
be estimated by calculating the total gravitational \textbf{specific}
force  in its surrounding few parsecs. Following Sanders
(\cite{sad98}) and assuming a core radius of 0.085 pc and the mass of
the SMBH to be 4.5$\times$10${^{6}}$ \s\,, it becomes clear that the
force from the SMBH and the surrounding star cluster is significant at
distances less than about 0.25 pc ($\sim$ 6 \as), i.e. in the region
of the \ohs\ head. It decreases by more than one order of magnitude at
distances greater than about 0.7 pc where the mid and tail parts of the
OH-streamer reside. Thus, the gravitational pull of the SMBH and the 
surrounding star cluster is strong enough to affect the kinematics 
of the head, but not the mid or tail.

In a recent study, Yusef-Zadeh et al. (\cite{yus13}) identified
blobs of weak radio emission at 8.4 GHz located on a line through the
position of \sgrastar\ at a PA of $\sim30\degr$ E of N. Three of the
blobs, b), c), and d), appear to be accommodated inside of the \ohs\
between the head and mid maxima. Furthermore, CN
$J=2-1$ emission is observed at the position of blob d) (RA, Dec) =
17$^{\rm h}$45$^{\rm m}$38\fs8, $-29\degr$$00'$40\farcs. We note that
this position also coincides with the region of intersection between
the eastern part of the +80 \kms\ cloud and the mid part of the \ohs,
i.e. where a steep intensity gradient is seen in
Fig. \ref{OHSCND802000_inv.pdf}.   

The +80 \kms\ cloud interacts with the \ohs\ and the CND,
while the cloud is not a part of the CND itself. The southern part of
the cloud interacts with the \ohs\ Mid and Tail parts, as seen both in
the map- and position-velocity planes. Interaction between the \ohs\
and the +80 \kms\ cloud in this region may also be indicated by the
striking increase in the value of $\Delta V_{\rm FWHM}$ of line profile
components here. Although shock-excited 1720-MHz OH maser sources were
not found in 1985 in this region of interaction (Karlsson et
al. \cite{kar03}), it may well be worthwhile repeating such a search at a later epoch. The velocity in the northern part of the +80 \kms\
cloud seems to be generally increasing to the north and to adapt to
the velocity of the CND NE lobe as observed in CN emission. 
This part of the cloud may, however, be a detached clump
of the +80 \kms\ cloud (Figs. B.1 and C.1).

Interaction between the +20 \kms\ cloud and the CND SW lobe (via the
SS) is consistent with the observation of Coil \& Ho (\cite{coi99})
that the northern tip of the +20 \kms\ cloud/(SS) interacts with (feeds)
the CND SW lobe. Additionally, Karlsson et al. (\cite{kar13}),
observing this region in the \ciso\ $J=2-1$ emission line, found that
the northern part of the +20\kms\ cloud/SS has an extension which
bends abruptly from the northwest to the northeast, pointing in towards the
\ohs\ tail and CND SW lobe (see their Fig. 11). The presence of shock
activity in this common region of the OH-streamer, the +20 and +80
\kms\ clouds, and the CND SW is furthermore supported
by recent $Odin$ observations of an unusually wide \htio\
positive-velocity absorption line. The abundance ratio of \hto\ with
respect to $\rm{H}_{2}$ was found to be approximately $1.4 \times
10^{-6}$, a high value indicative of the presence of strong
shocks desorbing water from dust grains in this region (Karlsson et al. \cite{kar13}, and Appendix
F). In Fig. \ref{Martin Fig 9 - 2 May.jpg}, we have tentatively indicated the
region of interaction (in the map plane) between the +20 and +80 \kms\
clouds and the \ohs.  

A new result in our data is that of a possible link between the near
side of the EMR and the CND SW lobe (Figs. \ref{decvellarge_2000.pdf}
and \ref{ravellarge_2.pdf}). Is the EMR a link in the process
of transporting material from the 100 pc scale to the 10 pc scale of the CND via molecular clouds (Fathi et al. \cite{fat06})? Incidentally, this region also coincides with the
rotating ring model of the CND where the velocity is about $-$115
\kms\  with a velocity width of 40 \kms\ (Mart\'in et al. \cite{mar12}). 
Furthermore, Emsellem et al. (\cite{ems14}) have
studied the interplay between a galactic bar and an SMBH by simulating
a Milky Way-like galaxy. They found that gas is focused on (Lindblad)
resonances into elongated ring-like structures and subsequently
connects to the SMBH via mini-spirals inside of a few tens of
pc. Ultimately, the gas accreted in the vicinity of the SMBH creates a
series of winding tails of gas, supporting our results in this
work. In particular, the \ohs\  resembles the features presented in
Emsellem et al. (\cite{ems14}), which suggest that 
it could have been triggered by tidal-like forces owing to the
interplay between the gravitational potential of the Galactic bar and
that of the SMBH at the centre of the Galaxy.  

\section{Conclusions}

We have presented an analysis of certain features in our concatenated data
base of VLA BnA and DnC observations of 1665 and 1667 MHz OH
absorption towards the Sgr A Complex at the GC. Our data have an
angular resolution of about 6\as\ and a velocity resolution of about 9
\kms, and the full data set has been made available to the scientific
community using the CDS \footnote{http://vizier.u-strasbg.fr/viz-bin/VizieR}. 

Our investigation of the properties and kinematics of the \ohs\ and
the +80 \kms\ cloud in OH absorption indicated that the \ohs\ is an
object inside of the CND and interacting with the +80 \kms\ cloud,
the CND, and possibly with the strong gravitational field from the
SMBH and the surrounding star cluster. We also ouind indications of
interaction between the +80 \kms\ cloud, or a detached clump of it, and the
northeastern lobe of the CND via the CND NE extension.  

We interpreted those kinematical and morphological links as indications
that gas clumps were disrupted from the SW lobe of the CND and may
have produced the \ohs\ and the +80 \kms\ cloud. At least the \ohs\
seems to be feeding material radially inwards, inside of the CND. For
the +80 \kms\ cloud the kinematics have still not been resolved. Although
the projected image of this cloud may suggest that it is a part of the
CND, the position-velocity diagrams clearly dispute this.  

Our conclusions are that the \ohs\ head is located in front of
\sgrastar\ at velocities between about 15 and 50 \kms\  and partly behind at
velocities between 59 and 67 \kms\  and that the \ohs\ head represents
a part of an inflow of gas from the CND region. In the mid and tail
parts, as well as in the head, we note a negative velocity gradient 
from the +80 \kms\ cloud and towards the GC, where one would expect 
increasing velocity inwards from the CND. If the \ohs\ and the +80 \kms\ cloud
are located on the ``far" side of the CND and behind \sgrawest, a flow
from ``behind" towards the centre would display such a negative velocity gradient.

Moreover, a detailed analysis of the position-velocity diagrams
revealed a possible link between the near side of the EMR and the CND
SW lobe. In this analysis we also found OH absorption against all four of the
compact \HII\ regions A - D, east of Sgr A East, both in the 1665 and
1667 MHz transitions.  

Further progress on the matter of relative locations and relations
between the \ohs, the +80 \kms\ cloud, and the surrounding molecular
clouds and \sgrastar would be gained from observations of OH with
still higher angular and velocity resolutions and from searches for those
objects in other species. Furthermore, a state-of-the-art 3D modelling
would be highly beneficial to reveal the relative locations of
molecular clouds in the 10 pc region surrounding the GC. 

\begin{acknowledgements}

The authors acknowledge the open policy for the use of NRAOs VLA. The
National Radio Astronomy Observatory (NRAO) is operated by Associated
Universities Inc., under cooperative agreement with the National
Science Foundation. Kambiz Fathi acknowledges support from the Swedish Royal 
Academy of Sciences' Crafoord Foundation. Furthermore, the extensive support 
from the Swedish National Space Board (Rymdstyrelsen) is gratefully
acknowledged by Aage Sandqvist. We also acknowledge with gratitude the
support from Miller Goss with concatenation of the observations with
the VLA in its BnA and DnC configurations. 

\end{acknowledgements}

\clearpage

\appendix

\section{OH absorption}

\begin{figure}[]
\includegraphics[angle=0, width=.49\textwidth]{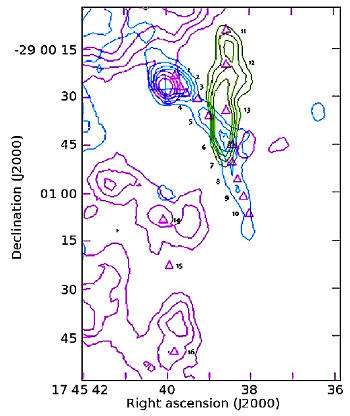}
\caption{Locations of the OH absorption profiles, labelled by
  triangles. The magenta contours 
  indicate the \ohs, the olive-coloured contours depict the +80
  cloud, and the purple contours delineate parts of the +20 \kms\
  cloud, at 50, 85, and 32 \kms, respectively. The lowest contour level
  is at $3.5\sigma$ (90 mJy/beam), and the contour interval is $1\sigma$
  spacing. Position numbers 1 - 10 belong to the \ohs, 
  11 - 13 are in the +80 \kms\ cloud, and 14 - 16 are inside the
  +20 \kms\ cloud.}  
\label{posnum.pdf}
\end{figure}

Figure \ref{posnum.pdf} is a map that indicates the positions where
  the profiles in Fig. A.2 were
  produced using the 1667-MHz OH absorption data cube. Positions 1 - 4
  are located in the \ohs\ head, 5 - 7 in 
  the mid, and 8 - 10 in the tail part. Positions 11 - 13 are in the
  +80 \kms\ cloud, and positions 14 - 16 are in the northern part of the
  +20 \kms\ cloud. The OH parameters, obtained from a Gaussian analysis
of these profiles, are presented in Table \ref{table:paramsum}.

Figure \ref{ravellarge_2.pdf} is a position-velocity diagram of
  the OH absorption 
  at 1667 MHz of the inner $7\am\times 7\am\ $ region of the GC. It is
  a visualisation through the entire data cube seen face-on from the
  right ascension-velocity side. The angular
  resolution is $7\as\ $, and the velocity resolution is 8.8 \kms. The
  prominent features are labelled in the figure. We note the bridge
  between the EMR near side and the CND SW lobe. 

\begin{figure}
\begin{flushleft}
\includegraphics[angle=0, width=.49\textwidth]{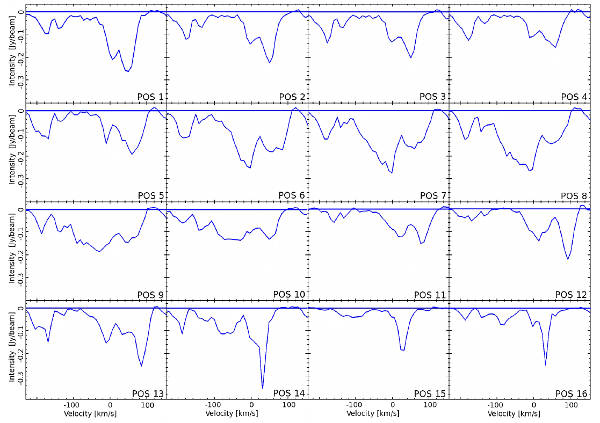}
\caption{1667-MHz OH absorption line profiles at the positions shown
  in Fig. \ref{posnum.pdf}. The angular resolution is
  $7\as \times 7\as$, and the velocity resolution is 8.8 \kms.}
\label{FigA2_16 profiles.jpg}
\end{flushleft}
\end{figure}

\begin{table*}
\caption{Parameters obtained from the Gaussian decomposition of the profiles at Positions 1 - 16  in Fig. A.1. $^{a}$ Intensities  $>$$3\sigma$, $^{b}$ ($V_{\mathrm{peak\ abs.}}$) = velocity at maximum absorption, and $^{c}$ ($\Delta V_{\rm FWHM}$) = linewidth. }
\centering\tabcolsep=2pt
\scalebox{1}{
\begin{tabular}{lcccccccl}
\hline\noalign{\smallskip}
Pos.\#\ & 
PA & 
$D$ fr. \sgrastar\ & 
$V_{\mathrm{peakabs.}}$\tablefootmark{b} & 
$\Delta T_{A}(1667)$, $\Delta T_{A}(1665)$\tablefootmark{a} &
$\Delta V_{\rm FWHM}$\tablefootmark{c}& 
${\Delta T_{A}(1667)}\over {\Delta T_{A}(1665)}$&  
$\tau_{1667}$& 
$N$(OH)/$T_\mathrm{ex}$ \\[0.5 ex]
\ & 
(deg)& 
(\as, pc) &
(\kms) & 
(Jy/beam)&
(\kms) & 
\ & 
\ & 
(cm$^{-2} \mathrm{K}^{-1}$)  \\  \hline\hline
1 \ohs\ head	& 315	& 5.1, 0.20	& 50	&$-$0.256, $-$0.173	& 33	&1.48	&1.0(+0.5/$-$0.4)  & $7.3\times 10{^{15}}$ \\   
2 		& 277	& 6.3, 0.25	& 49	&$-$0.216, $-$0.150	& 40	&1.44	&1.0(+0.7/$-$0.2)  & $8.7\times 10{^{15}}$ \\
3		& 263	& 7.6, 0.30	& 47	&$-$0.194, $-$0.125	& 35	&1.55	&0.9(+0.3/$-$0.4)  & $7.3\times 10{^{15}}$ \\
4 		& 256	& 10.1, 0.40	& 56	&$-$0.150, $-$0.094	& 35	&1.60	&0.6(+0.4/$-$0.4)  & $4.7\times 10{^{16}}$ \\  \hline\smallskip
5 \ohs\ mid 	& 239	& 16.1, 0.64	&57 	&$-$0.194, $-$0.131	& 35 	&1.48	&1.2(+0.3/$-$0.6)  & $ 9.3\times 10{^{15}}$\\
\		& 239	& 16.1, 0.64	&84 	&$-$0.127, $-$0.108	& 21 	&1.18	&2.9(+1.6/$-$0.8)  & $1.3\times 10{^{16}}$ \\ 
6		& 230	& 27.5, 1.10	&52	&$-$0.183, $-$0.113 	& 26	&1.62	&0.5(+0.4/$-$0.3)  & $2.9\times 10{^{15}}$ \\
\ 		& 230	& 27.5, 1.10	&80	&$-$0.173, $-$0.115	& 40	&1.50	&1.0(+0.4/$-$0.4)  & $8.7\times 10{^{15}}$ \\
\ 		& 230	& 27.5, 1.10	&33	&$-$0.151, $-$0.111	& 38	&1.36	&1.6(+0.6/$-$0.5)  & $1.3\times 10{^{16}}$ \\
7		& 223	& 30.6, 1.22	&57	&$-$0.166, $-$0.102	& 50	&1.63	&0.5(+0.3/$-$0.3)  & $5.5\times 10{^{15}}$ \\  \hline\noalign{\smallskip}
8 \ohs\ tail	& 219	& 36.1, 1.44	& 54	&$-$0.146, $-$0.101	&45	&1.45	&1.2(+0.4/$-$0.4)   & $ 1.2\times 10{^{16}}$ \\ 
9 		& 217	& 41.4, 1.66	& 45	&$-$0.142, $-$0.088	&48	&1.61	&0.6(+0.3/$-$0.4)   & $ 6.5\times 10{^{15}}$ \\
\ 		& 217	& 41.4, 1.66	& 77	&$-$0.109, $-$0.085	&24	&1.28	&2.1(+0.8/$-$0.6)   & $ 1.1\times 10{^{16}}$ \\
10 		& 214	& 46.7, 1.87	& 50	&$-$0.130, $-$0.085	&43	&1.53	&0.8(+0.4/$-$0.3)   & $ 8.0\times 10{^{15}}$ \\ \hline\noalign{\smallskip}
11 +80 \kms\ cl.& 318	& 28.5, 1.14	& 80	&$-$0.147, $-$0.132	& 38	&1.11	&3.9(N/A$/-$1.3)    & $3.3\times 10{^{16}}$ \\ 
\		& 318	& 28.5, 1.14	& 19	&$-$0.120, $-$0.076	& 50	&1.58	&0.7(+0.3/$-$0.4)   & $8.0\times 10{^{15}}$ \\
12  		& 308	& 24.7, 0.99 	& 92	&$-$0.224, $-$0.175	& 33	&1.28	&2.1(+0.8/$-$0.6)   & $1.7\times 10{^{16}}$ \\
\ 		& 308	& 24.7, 0.99 	& 13	&$-$0.142, $-$0.094	& 13	&1.51	&0.9(+0.4/$-$0.4)   & $1.0\times 10{^{15}}$ \\
13 		& 280	& 19.6, 0.78 	& 84	&$-$0.253, $-$0.209	& 33	&1.21	&2.6(+1.3/$-$0.7)   & $2.9\times 10{^{15}}$ \\
\ 		& 280	& 19.6, 0.78 	& 38	&$-$0.113, $-$0.087	& 44	&1.30	&1.9(+0.8/$-$0.5)   & $2.1\times 10{^{15}}$ \\ \hline\noalign{\smallskip}
14 +20 \kms\ cl.& 181	&41.0, 1.64	& 34	&$-$0.368, $-$0.269	& 16	&1.37	& 1.5(+0.7/$-$0.4)  & $5.4\times 10{^{15}}$ \\
15		& 182	&55.6, 2.22	& 28	&$-$0.184, $-$0.138	& 27	&1.33	& 1.9(+0.6/$-$0.7)  & $2.1\times 10{^{15}}$ \\  
16		& 184	&81.6, 3.26	& 32	&$-$0.244, $-$0.197	& 17	&1.24	& 2.4(+1.0/$-$0.7)  & $9.3\times 10{^{15}}$ \\
\noalign{\smallskip}\hline\\
\end{tabular}
\label{table:paramsum}}
\end{table*}

\begin{figure}
\begin{center}
\includegraphics[angle=0, width=.40\textwidth]{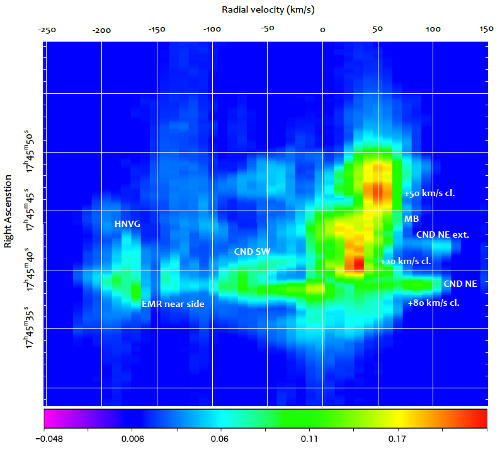}
\caption{Position-velocity diagram (RA, Vel) of OH absorption at 1667
  MHz. This is a visualisation through the entire data cube
  as seen from the right ascension-velocity side. The 1665 MHz data overlap
  at velocities higher than about 160 \kms, see Fig. 3 in Sandqvist
  (\cite{san73}).The wedge scale indicates the OH absorption in
  Jy/beam. (``HNVG'' stands for high negative velocity gas, ``MB'' the
  molecular belt.)}
\label{ravellarge_2.pdf}
\end{center}
\end{figure}

\section{1667 MHz OH-absorption}

\begin{figure*}[p]
\begin{center}
\includegraphics[angle=0, width=.99\textwidth]{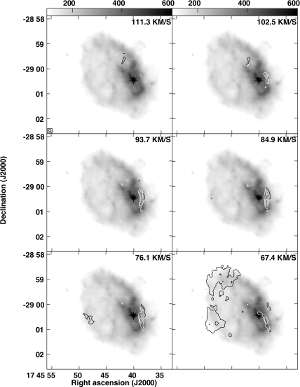}
\caption{1667-MHz OH-absorption at $111>V_{\rm LSR}>67$ \kms. The lowest contour level is 50 mJy/beam ($\sim2$$\sigma$) and the contour spacing is also $\sim2$$\sigma$. The wedge scale is in mJy/beam. The position of Sgr A* is marked with a plus sign.}
\label{1667_1.pdf}
\end{center}
\end{figure*}

\begin{figure*}[p]
\begin{center}
\includegraphics[angle=0, width=.99\textwidth]{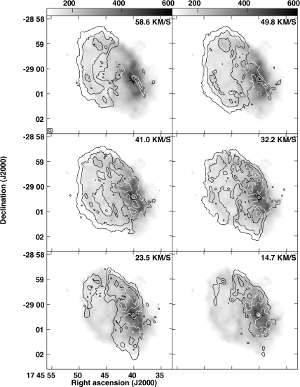}
\caption{1667-MHz OH-absorption at $59>V_{\rm LSR}>15$ \kms. The lowest contour level is 50 mJy/beam ($\sim2$$\sigma$), and the contour spacing is also $\sim2$$\sigma$. The wedge scale is in mJy/beam. The position of Sgr A* is labelled with a plus sign.}
\label{1667_2.pdf}
\end{center}
\end{figure*}

\begin{figure*}[p]
\begin{center}
\includegraphics[angle=0, width=.99\textwidth]{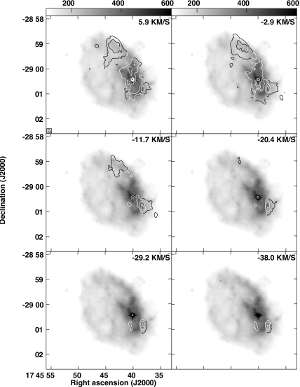}
\caption{1667-MHz OH-absorption at $6>V_{\rm LSR}>-38.0$ \kms. The lowest contour level is 50 mJy/beam ($\sim2$$\sigma$), and the contour spacing is also $\sim2$$\sigma$. The wedge scale is in mJy/beam. The position of Sgr A* is labelled with a plus sign.}
\label{1667_3.pdf}
\end{center}
\end{figure*}

\begin{figure*}[p]
\begin{center}
\includegraphics[angle=0, width=.99\textwidth]{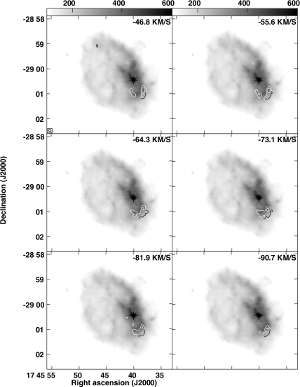}
\caption{1667-MHz OH-absorption at $-47>V_{\rm LSR}>-91$ \kms. The lowest contour level is 50 mJy/beam ($\sim2$$\sigma$), and the contour spacing is also $\sim2$$\sigma$. The wedge scale is in mJy/beam. The position of Sgr A* is labelled with a plus sign.}
\label{1667_4.pdf}
\end{center}
\end{figure*}

\begin{figure*}[p]
\begin{center}
\includegraphics[angle=0, width=.99\textwidth]{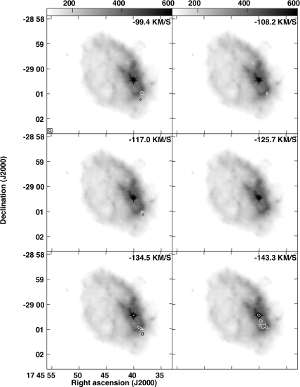}
\caption{1667-MHz OH-absorption at $-99>V_{\rm LSR}>-143$ \kms. The lowest contour level is 50 mJy/beam ($\sim2$$\sigma$), and the contour spacing is also $\sim2$$\sigma$. The wedge scale is in mJy/beam. The position of Sgr A* is labelled with a plus sign.}
\label{1667_5.pdf}
\end{center}
\end{figure*}

\begin{figure*}[p]
\begin{center}
\includegraphics[angle=0, width=.99\textwidth]{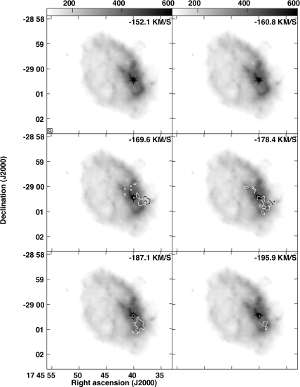}
\caption{1667-MHz OH-absorption at $-152>V_{\rm LSR}>-196$ \kms. The lowest contour level is 50 mJy/beam ($\sim2$$\sigma$), and the contour spacing is also $\sim2$$\sigma$. The wedge scale is  in mJy/beam. The position of Sgr A* is labelled with a plus sign.}
\label{1667_6.pdf}
\end{center}
\end{figure*}

\section{$-T_{\rm{L}}/T_{\rm{C}}$ at 1667 MHz}

\begin{figure*}[p]
\begin{center}
\includegraphics[angle=0, width=.99\textwidth]{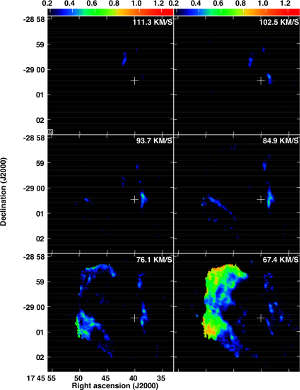}
\caption{$-T_{\rm{L}}/T_{\rm{C}}$ at 1667 MHz $111>V_{\rm LSR}>67$ \kms. The position of Sgr A* is shown with a plus sign.}
\label{TLTC1C.pdf}
\end{center}
\end{figure*}

\begin{figure*}[p]
\begin{center}
\includegraphics[angle=0, width=.99\textwidth]{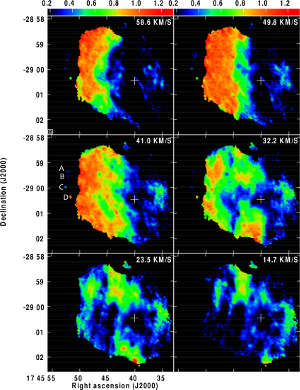}
\caption{$-T_{\rm{L}}/T_{\rm{C}}$ at 1667 MHz $59>V_{\rm LSR}>15$ \kms. The position of Sgr A* is shown with a plus sign, and the four Compact \HII\ regions are marked by letters A-D.} 
\label{TLTC2C.pdf}
\end{center}
\end{figure*}

\begin{figure*}[p]
\begin{center}
\includegraphics[angle=0, width=.99\textwidth]{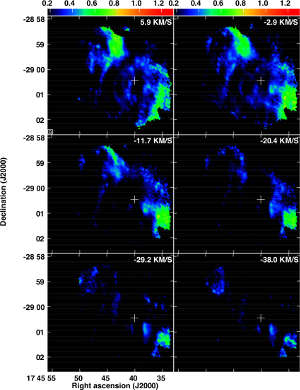}
\caption{$-T_{\rm{L}}/T_{\rm{C}}$ at 1667 MHz $6>V_{\rm LSR}>-38$ \kms. The position of Sgr A* is shown with a plus sign.}
\label{TLTC3C.pdf}
\end{center}
\end{figure*}

\begin{figure*}[p]
\begin{center}
\includegraphics[angle=0, width=.99\textwidth]{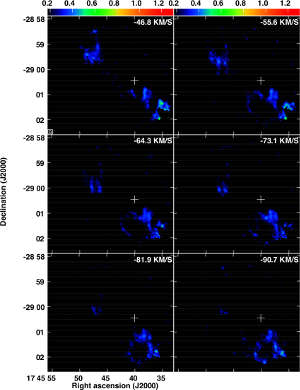}
\caption{$-T_{\rm{L}}/T_{\rm{C}}$ at 1667 MHz $-47>V_{\rm LSR}>-91$ \kms. The position of Sgr A* is shown with a plus sign.}
\label{TLTC4C.pdf}
\end{center}
\end{figure*}

\begin{figure*}[p]
\begin{center}
\includegraphics[angle=0, width=.99\textwidth]{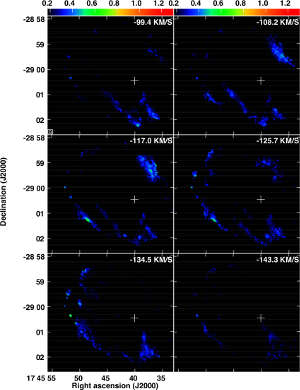}
\caption{$-T_{\rm{L}}/T_{\rm{C}}$ at 1667 MHz $-99>V_{\rm LSR}>-143$ \kms.. The position of Sgr A* is shown with a plus sign.}
\label{TLTC5C.pdf}
\end{center}
\end{figure*}

\begin{figure*}[p]
\begin{center}
\centering
\includegraphics[angle=0, width=.99\textwidth]{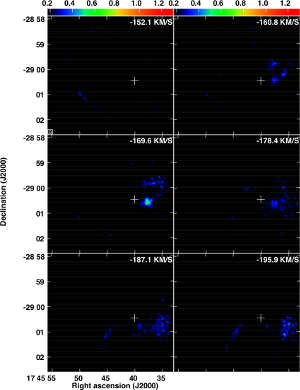}
\caption{$-T_{\rm{L}}/T_{\rm{C}}$ at 1667 MHz $-152>V_{\rm LSR}>-196$ \kms. The position of Sgr A* is shown with a plus sign.}
\label{TLTC6C.pdf}
\end{center}
\end{figure*}

\section{1665 MHz OH-absorption}

\begin{figure*}[p]
\begin{center}
\includegraphics[angle=0, width=.99\textwidth]{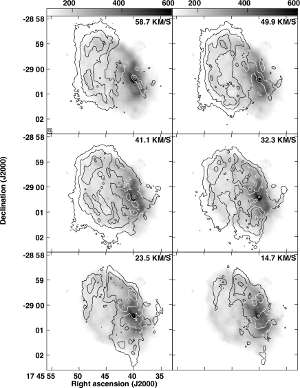}
\caption{1665-MHz OH-absorption at $59>V_{\rm LSR}>15$ \kms. The lowest contour level is 50 mJy/beam ($\sim2$$\sigma$), and the contour spacing is also $\sim2$$\sigma$. The wedge scale is in mJy/beam. The position of Sgr A* is labelled with a plus sign.}\label{1665_2.pdf}
\end{center}
\end{figure*}

\section{$-T_{\rm{L}}/T_{\rm{C}}$ at 1665 MHz}

\begin{figure*}[p]
\begin{center}
\includegraphics[angle=0, width=.99\textwidth]{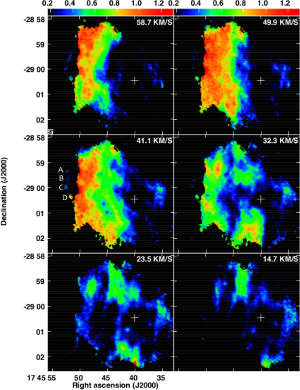}
\caption{$-T_{\rm{L}}/T_{\rm{C}}$ at 1665 MHz at $59>V_{\rm LSR}>15$ \kms. The position of Sgr A* is labelled with a plus sign, and the four Compact \HII\ regions are labelled by letters A-D.}
\label{TAU65_3.pdf}
\end{center}
\end{figure*}

\section{H$_2^{18}$O in the \ohs/+80 \kms\ cloud/CND SW shock region}

\begin{figure}[h]
\includegraphics[angle=0, width=.49\textwidth]{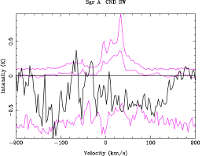}
  \caption{\htio\ (black line), \htmo\ (lower magenta line), \ciso\
    $J=1-0$ (middle magenta line), and \ciso\ $J=2-1$ (upper magenta
    line) profiles towards the SW position in the CND, coinciding with
    the position of interaction between the OH Streamer, the +80
    \kms\ cloud and the SS/+20 \kms\ cloud. The \htio\ antenna
    temperature scale has been multiplied by a factor of 10. The
    \htmo\ antenna temperature scale has been lowered by 0.7 K for
    clarity. The intensity scales of the two \ciso\ profiles are in
    units of brightness temperature, the $J=2-1$ profile having been
    raised by 0.1 K for clarity. The channel resolution is 3 \kms\ for
    all the profiles.}  
  \label{cndswh20c18ocf.pdf}
\end{figure}

To further investigate the proposed shock region towards the
SW part of the CND, we have used the $Odin$ satellite to
observe the water isotope, \htio, which is tracer of shock (or
strong UV) regions (Karlsson et al. \cite{kar13}). The observations
were performed towards the position (RA, Dec) of $17^h45^m39\farcs7,
-29\degr01\arcmin18\arcsec$ (J2000.0) during April 2013 and April
2014. The total ON-source integration time of the combined data was 62.5
hours. An \htmo\ profile was also obtained towards the same position
in February 2013, with a total ON-source integration time of 9.4
hours. $Odin$'s HPBW at the \hto\ frequencies is
2\farcm1. The profiles are shown in Fig. \ref{cndswh20c18ocf.pdf}, together with \ciso\ $J=1-0$
and $2-1$ profiles obtained with SEST. We refer to Karlsson et
al. (\cite{kar13}) for a detailed description of $Odin$ and SEST
observations and analysis of the \sgracomp\ region. Here we simply
present our new results and their interpretation.  

The \htio\ profile shows a remarkably wide absorption component at
positive velocities, in addition to the well-known expanding molecular ring (EMR) and 3-kpc Arm
features near $-130$ and $-50$ \kms, respectively. This
positive-velocity region covers the velocity range of the interacting
components discussed in the main part of the paper, {\it viz.} the OH-Streamer, the +80 \kms\
cloud and the +20 \kms\ cloud/SS. We obtain a total column density of
the order of $N$(\htio) $\sim 2.2 \times 10^{14}$ cm$^{-2}$, which
corresponds to $N$(\htmo) $\sim 5.5 \times 10^{16}$, assuming a
\htmo/\htio\ abundance ratio of 250 in the inner Galactic centre (Wilson
\& Rood \cite {wil94}). From our two \ciso\ lines we obtain, in the same region
and velocity interval, a molecular hydrogen column density of $N$(H$_2$)
$\approx 4.0 \times 10^{22}$ cm$^{-2}$, assuming a \ciso/H$_2$ abundance ratio of
$2 \times 10^{-7}$ (Goldsmith \cite{gol99}). These values result in an
abundance ratio [o-\hto/H$_2$] = $X$(o-\hto) $\sim 1.4 \times
10^{-6}$. Such a high abundance ratio of \hto\ is comparable to what
is found in, for example, the low-velocity outflow region of Orion (Persson et
al. \cite{per07}), where it has been interpreted as desorption of water
ice from dust grains due to shock effects.

\end{document}